\documentclass[preprint,nofootinbib]{revtex4}
\pdfoutput=1

\usepackage{graphicx}
\usepackage{amsmath}
\usepackage{amssymb}
\usepackage{ulem}
\usepackage{xcolor}
\graphicspath{{./figs/}}

\begin{document}

\title{Unphysical properties in a class of interacting dark energy models}

\author{R. von Marttens$^{1,2}$, H. A. Borges$^{3}$, S. Carneiro$^{3,4}$, J. S. Alcaniz$^{1}$, W. Zimdahl$^{4}$}
\affiliation{$^{1}$Observat\'orio Nacional, 20921-400, Rio de Janeiro, RJ, Brasil\\$^{2}$D\'epartment de Physique Th\'eorique and Centre for Astroparticle Physics, Universit\'e de Gen\`eve, Quai E. Ansermet 24, CH-1211 Gen\`eve 4, Switzerland\\ $^3$Instituto de F\'{\i}sica, Universidade Federal da Bahia, 40210-340, Salvador, BA, Brasil\\$^4$PPGCosmo, CCE, Universidade Federal do Esp\'irito Santo, 29075-910, Vit\'oria, ES, Brasil}

\date{\today}

\begin{abstract}
Models with non-gravitational interactions between the dark matter and dark energy components are an alternative to the standard cosmological scenario. These models are characterized by an interaction term, and a frequently used parameterization is $Q = 3\xi H \rho_{x}$, where $H$ is the Hubble parameter and
$\rho_{x}$ is the dark energy density. Although recent analyses have reported that  this particular scenario provides a potential solution to the $H_{0}$ and $\sigma_{8}$ tensions for negative values of the interaction parameter $\xi$,   we show here that such an interval of values of $\xi$ leads to a violation of the Weak Energy Condition for the dark matter density,  
which is accompanied by unphysical instabilities of matter perturbations. Using current observational data we also show that the inclusion of the physical prior $\xi \geq 0$ in the statistical analysis alters the parameter selection for this model and discards it as a solution for the $H_{0}$-tension problem. 
\end{abstract}

\maketitle

\section{Introduction}

As it is well known, there is no known fundamental principle that prevents a non-minimal coupling between the energy components of the cosmological dark sector.  Such a possibility has in fact been explored since the eighties as an alternative to the standard cosmology (see e.g. ~\cite{Ozer:1985ws,Ozer:1985wr,Freese:1986dd}), with its theoretical and observational consequences being of great interest nowadays~\cite{Valiviita:2008iv,Carneiro:2014uua,Borges:2017jvi,vonMarttens:2018iav,Benetti:2019lxu,Xia:2016vnp,DiValentino:2017iww,Kumar:2017dnp,Yang:2018uae,Yang:2019uzo,Kumar:2019wfs,DiValentino:2019dzu,Pan:2019jqh,Martinelli:2019dau,DiValentino:2019jae,Yang:2018euj,Pan:2019gop,Yang:2017ccc,Pan:2020bur,vonMarttens:2019ixw}. However, in the absence of a natural guidance from fundamental physics on the coupling term, a number of  phenomenological models have been proposed and their cosmological consequences investigated in light of the current observational data -- we refer the reader to \cite{Cid:2018ugy} for a recent comparative analysis of different classes of interacting models.

In particular, models in which the coupling or interacting term $Q$ is proportional to the dark energy (DE) density $\rho_{x}$ \cite{Zimdahl:2001ar,Chimento:2003iea,Setare:2006wh,Sadjadi:2006qp}, $Q = 3\xi H \rho_{x}$, where $H$ and $\xi$ are the Hubble and interaction parameters respectively, have become popular in the recent years~(see e.g. \cite{DiValentino:2019ffd} and references therein). However, as will be shown in this paper, this class of models shows unphysical behavior for the interval of values of its parameters currently constrained by observational data \cite{DiValentino:2019ffd}. In particular, the model predicts that the pressureless matter density will eventually become negative, violating the Weak Energy Condition (WEC), with further consequences for the evolution of the dark matter and baryon density perturbations.

\section{Background dynamics}

Let us first consider 
the balance equations of the model
\begin{eqnarray}
\dot{\rho}_b + 3H \rho_b &=& 0, \label{barions} \\
\dot{\rho}_{dm} + 3H \rho_{dm} &=& 3 \xi H \rho_x, \label{materia_escura}\\
\dot{\rho}_x + 3H \rho_x (1+\omega_x) &=& -3 \xi H \rho_x, \label{DE}
\end{eqnarray}
where $\rho_{dm}$ and $\rho_{b}$ are respectively the densities of dark matter (DM) and conserved baryons, 
$\omega_x < 0$ is the DE equation-of-state (EoS) parameter, and a dot means derivative with respect to   (w.r.t.) cosmological time. Summing up equations (\ref{barions}) and (\ref{materia_escura}) gives ($ \rho_{m} = \rho_{dm}+ \rho_b$)
\begin{equation}
\dot{\rho}_{m} + 3H \rho_{m} = 3 \xi H \rho_x \label{materia}
\end{equation}
for the total pressureless matter.
The general solutions of (\ref{DE})-(\ref{materia}) are
\begin{eqnarray} \label{4}
\rho_m &=& C a^{-3} - \left( \frac{\xi \rho_{x0}}{\xi+\omega_x} \right) a^{-3(\xi+\omega_x+1)}\;,\\
\label{3}
\rho_x &=& \rho_{x0}\, a^{-3(\xi+\omega_x+1)}\;.
\end{eqnarray}
In the spatially flat case, $C$ and $\rho_{x0}$ obey the additional constraint $\rho_{m0} + \rho_{x0} = 3H_0^2$, where a subscript $0$ denotes the value of the corresponding quantity at present time. That is,
\begin{equation}\label{C}
C = 3H_0^2 \left( 1 - \frac{\omega_x \Omega_{x0}}{\xi+\omega_x} \right),
\end{equation}
where $\Omega_{x0} = \rho_{x0}/(3H_0^2)$ is the present DE density parameter. 

In order to assess the conditions that lead to the WEC violation, it is convenient to split the analysis into two parts, namely, the past and future WEC. From \eqref{4}, it is possible to show that the matter energy density becomes negative at
\begin{equation} \label{a}
a = \left[ \frac{\xi \rho_{x0}}{(\xi+\omega_x)C} \right]^{\frac{1}{3(\xi + \omega_x)}}.
\end{equation}
Assuming that $w_{x}<0$, it is straightforward to obtain the conditions that will provide solutions for Eq.~\eqref{a} in the intervals $0<a<1$ (past) and $a>1$ (future). To ensure that the matter energy density does not assume negative values at early times, the interaction parameter must obey the relation 
\begin{equation} \label{prior1}
    \xi < |\omega_x|\Omega_{m0}\,.
\end{equation}
On the other hand, the second term on the right-hand side of equation (\ref{4}) is negative and it will eventually dominate $\rho_m$ in the future unless
\begin{equation} \label{prior2}
    \xi \geq 0\,,
\end{equation}
regardless of the  value of $w_{x}$. Eqs.~\eqref{prior1} and \eqref{prior2}, therefore, define the range where the model is physically well-defined for any time scale. In particular, the latter relation will be used in Sec.~\ref{paramselec} to define the WEC or physical prior when constraining the cosmological parameters within non-prohibited regions.

For example, if $\omega_x = -1$, $\Omega_{x0} = 0.7$ and $\xi = -0.1$, $\rho_m$ becomes negative when the scale factor is $a \approx 1.7$. In Fig. 1 we show the scale factor for WEC violation as a function of the interaction parameter $\xi$. Figure 2  shows the behavior of the Hubble function, as well as of the DE and matter densities, as functions of the scale factor. 
The violation of the WEC in this class of interacting models was also pointed out in \cite{Borges:2017jvi}.

\begin{figure*}
\begin{center}
\includegraphics[width=0.45\textwidth]{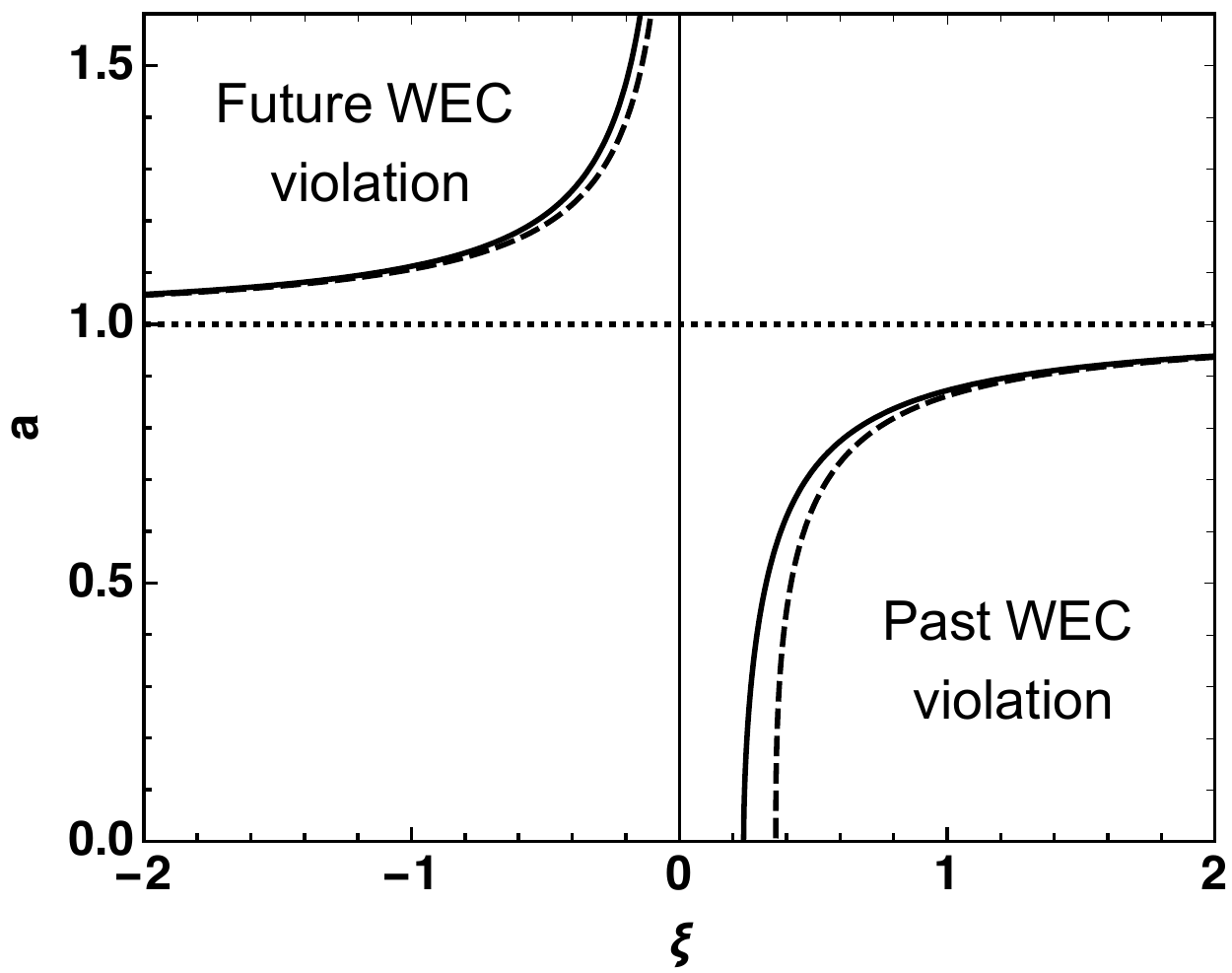}
 \end{center}
 \caption{Scale factor for WEC violation as a function of the interaction parameter $\xi$. The solid line corresponds to $w_{x}=-0.8$ and the dashed line corresponds to $w_{x}=-1.2$. The DE density parameter was fixed to $\Omega_{x0} = 0.7$. The result for $w_{x}=-1$ is in between solid and dashed lines.}
  \label{WEC}
 \end{figure*}

\section{Perturbations}

\begin{figure*}
\centerline{\includegraphics[width=0.3\textwidth]{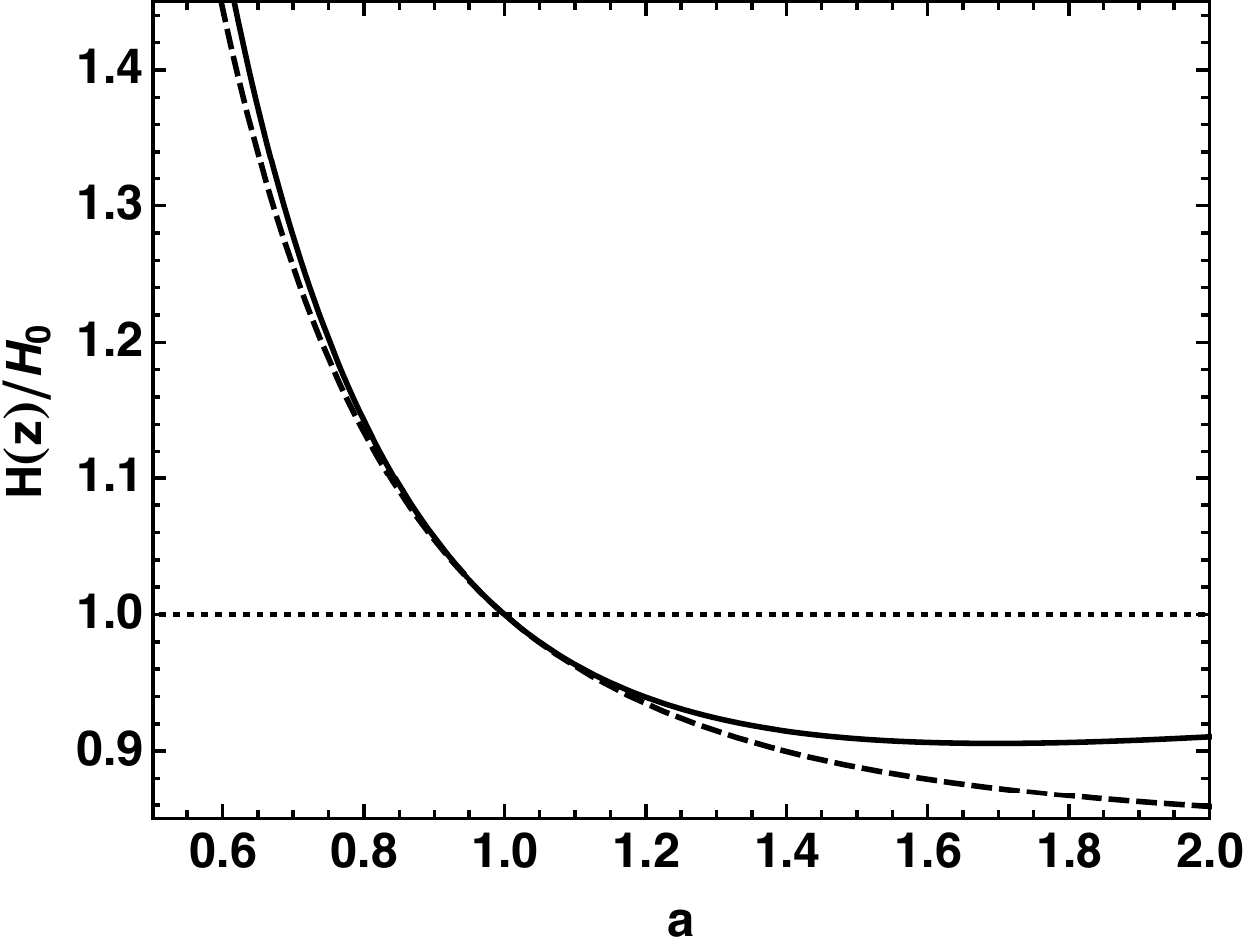} \hspace{.0in} \includegraphics[width=0.3\textwidth]{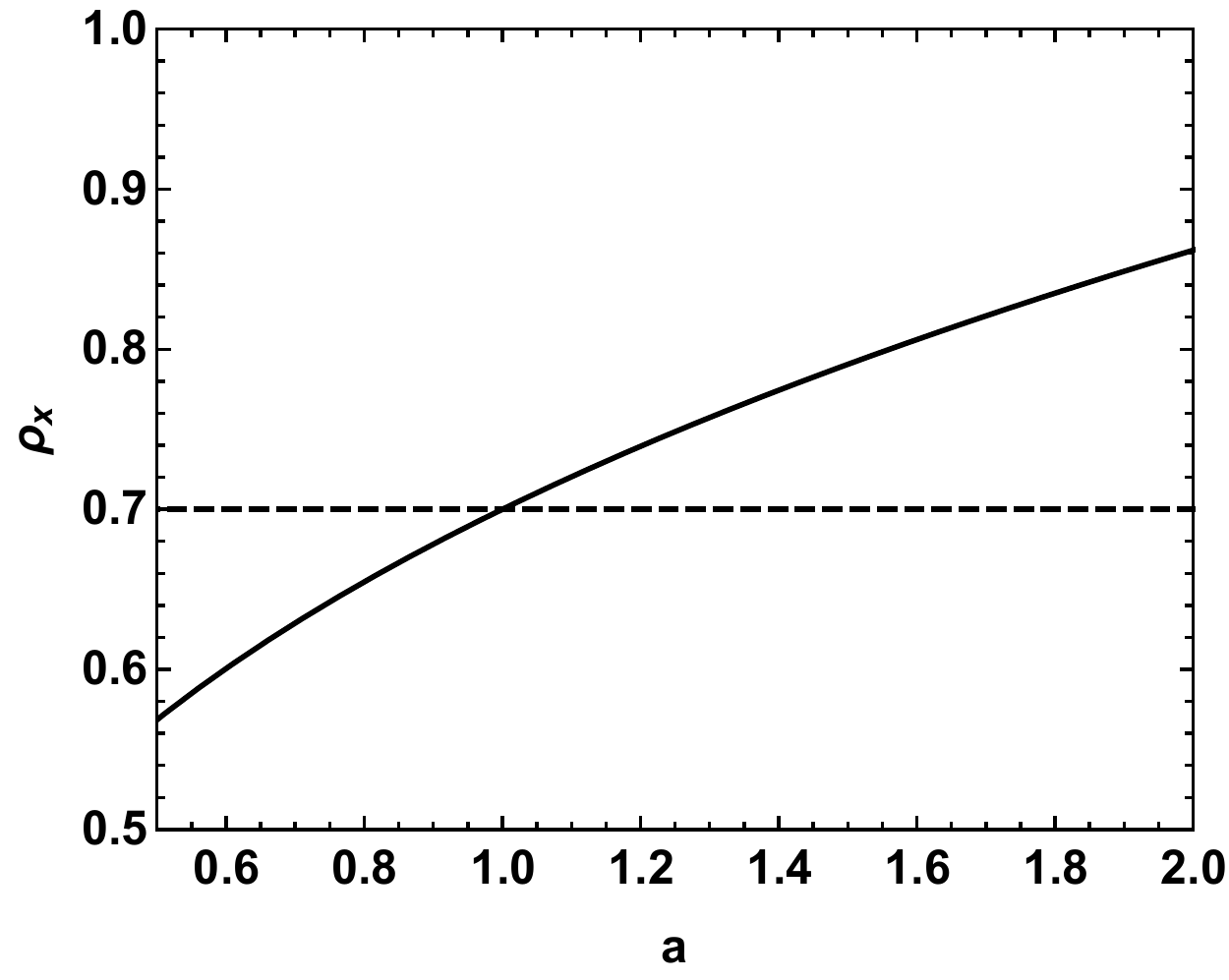} \hspace{.0in} \includegraphics[width=0.3\textwidth]{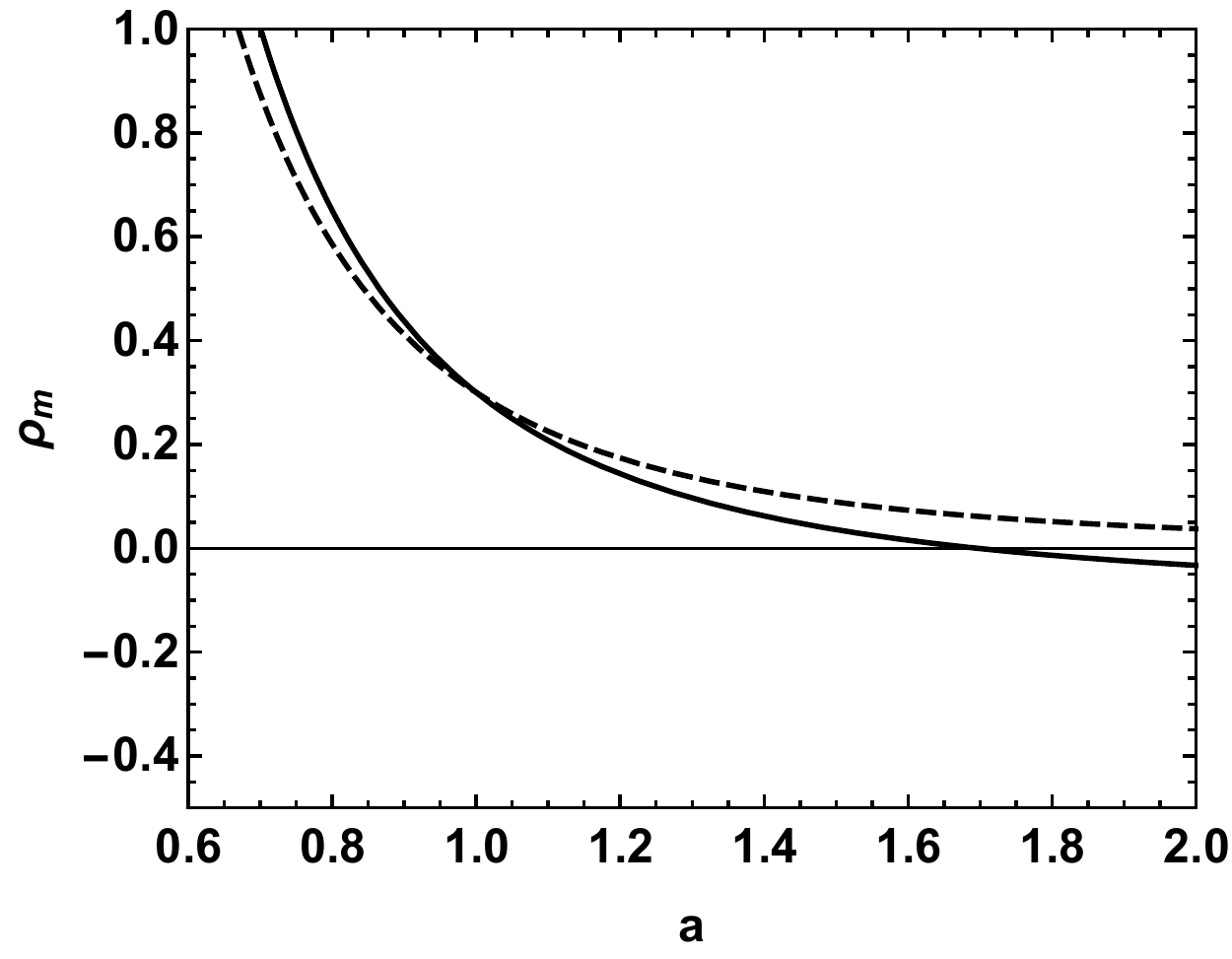}}
 \caption{Background solutions as functions of the scale factor. \textbf{Left panel:} Hubble function $H/H_0$. \textbf{Centre panel:} DE relative energy density. \textbf{Right panel:} Pressureless matter relative energy density. In all panels solid lines correspond to $\xi = -0.1$, $\omega_x = -1$ and $\Omega_{x0} = 0.7$ and dashed lines correspond to the $\Lambda$CDM model with the same $\Omega_{x0}$.}
  \label{H[a]}
 \end{figure*}

In a model with non-gravitational interaction in the dark sector, the conservation equations for conserved baryons, DM and DE are
\begin{equation}
\label{tres}
{T_{b}^{\mu\nu}}_{;\nu}=0,
\end{equation}
\begin{equation}
\label{quatro'}
{T_{dm}^{\mu\nu}}_{;\nu}=Q^{\mu},
\end{equation}
\begin{equation}\label{cinco}
{T_{x}^{\mu\nu}}_{;\nu}=-Q^{\mu}.
\end{equation}
Combining Eqs. (\ref{tres}) and (\ref{quatro'}) we obtain
\begin{equation}
\label{quatro}
{T_{m}^{\mu\nu}}_{;\nu}=Q^{\mu}.
\end{equation}
We can decompose the energy-momentum transfer $Q^{\mu}$ in directions parallel and orthogonal to the fluid $4$-velocity $u^{\mu}$,
\begin{equation}\label{seis}
Q^{\mu}=Qu^{\mu}+\bar Q^{\mu},
\end{equation}
with
\begin{equation}
u^{\mu}\bar Q_{\mu}=0.
\end{equation}
For comoving observers, $Q$ represents the energy transfer between the components, and $\bar{Q}^{\mu}$ represents the momentum transfer.

Assuming that both DE and pressureless matter are adiabatic perfect fluids, perturbing ($\ref{cinco}$) and ($\ref{quatro}$) in the longitudinal gauge, and assuming that there is no momentum transfer in the matter rest frame (i.e. matter follows geodesics), we find the balance and Poisson equations
\begin{equation}\label{thetad}
\theta'_{m}+\mathcal{H}\theta_{m}-k^2\Phi=0,
\end{equation}
\begin{equation} \label{thetax}
\theta'_x+\bigg[\frac{\omega'_x}{1+\omega_x}-\frac{aQ}{\rho_x}-\mathcal{H}(3\omega_x-1)\bigg]\theta_x-k^2\Phi=\frac{1}{1+\omega_x}\bigg[\frac{k^2}{\rho_x}\delta p_ x-a\frac{Q}{\rho_x}\theta_{m})\bigg],
\end{equation}
\begin{equation}\label{deltad}
\delta'_{m}-3\Phi'+\theta_{m}=-\frac{aQ}{\rho_{m}}\delta_{m}+ \frac{aQ}{\rho_{m}}\Phi+\frac{a\delta Q}{\rho_{m}},
\end{equation}
\begin{equation} \label{deltax}
\frac{[\delta\rho_x]'}{\rho_x}+3\mathcal{H}\bigg[\delta_x+\frac{\delta p_ x}{\rho_x}\bigg]-3(1+\omega_x)\Phi'+(1+\omega_x)\theta_x=-\frac{aQ}{\rho_x}\Phi-\frac{a\delta Q}{\rho_x},
\end{equation}
\begin{equation}\label{Poisson}
-\Phi=\frac{a^2}{2}\frac{\rho}{k^2}\delta+\frac{3a^2}{2}\bigg(\frac{\mathcal{H}^2}{k^2}\bigg)\bigg(\frac{\rho_m}{k^2}\bigg)\frac{\theta}{\mathcal{H}},
\end{equation}
where a prime means derivative w.r.t. conformal time, $\mathcal{H} = aH$, $\rho$ is the total energy density, $\delta =\delta \rho/\rho$, and $\theta$ is the fluid velocity potential.

Multiplying (\ref{thetax}) by $(1+\omega_x)$ and setting $\omega_x = -1$, we obtain
\begin{equation} \label{deltax2}
\delta \rho_x = -\frac{aQ\theta_{m}}{k^2}.
\end{equation}
Using (\ref{deltax2}) and (\ref{thetad}) in (\ref{deltax}),
\begin{equation} \label{deltaQ}
\delta Q = \frac{Q'\theta_{m}}{k^2}.
\end{equation}
The remaining equations lead to the Poisson and matter perturbation equations
\begin{equation}\label{thetad2}
\theta'_{m}+\mathcal{H}\theta_{m}-k^2\Phi=0,
\end{equation}
\begin{equation}\label{deltad2}
\delta'_{m}-3\Phi'+\theta_{m}=-\frac{aQ}{\rho_{m}} \left[ \delta_{m} - \frac{1}{k^2} \left( k^2 \Phi + \frac{Q'}{Q}\theta_{m} \right) \right],
\end{equation}
\begin{equation} \label{poisson2}
-k^2 \Phi=\frac{a^2}{2} \rho_{m} \delta_{m} - \left( \frac{a^3 Q}{2} - \frac{3a^2}{2}\mathcal{H}\rho_m \right)\frac{\theta_{m}}{k^2},
\end{equation}
where $\theta = \theta_{m}$ (the DE velocity remains undefined).
In the limit of small scales, Eqs. (\ref{thetad2})-(\ref{poisson2}) are reduced to
\begin{eqnarray}\label{matter1}
\theta'_{m}+\mathcal{H}\theta_{m}+\frac{a^2}{2} \rho_{m} \delta_{m}=0,\\
\label{matter2}
\delta'_{m}+\theta_{m}=-\frac{aQ}{\rho_{m}} \delta_{m},
\end{eqnarray}
whereas (\ref{deltax2}) and (\ref{deltaQ}) are negligible in this limit. For conserved baryons the balance equations are
\begin{eqnarray}\label{barions1}
\theta'_{b}+\mathcal{H}\theta_{b}+\frac{a^2}{2} \rho_{m} \delta_{m}=0,\\
\label{barions2}
\delta'_{b}+\theta_{b}=0.
\end{eqnarray}

There is no pressure term in equations (\ref{matter1})-(\ref{barions2}). Instabilities are caused by the background interaction term $Q = 3 \xi H \rho_x$ due to the violation of the  WEC discussed earlier.  
Some authors try to fix this issue by taking $\omega_x \neq -1$. In this case, the system of perturbation equations (\ref{thetad})-(\ref{Poisson}) does not close. Then an additional ansatz for $\delta Q$ is needed, but it should be consistent to a covariant perturbation of the background ansatz $Q = 3 \xi H \rho_x$. Another possibility is to consider interacting models with a non-adiabatic DE, with sound speed $c_s^2 = 1$, which on the other hand leads in general to large scale instabilities \cite{Valiviita:2008iv,Clemson:2011an} (see also \cite{Jackson:2009mz,Majerotto:2009np}). In any case, the weak energy condition is still violated by the background in the present model. 
\begin{figure*}
\begin{center}
\centerline{\includegraphics[width=0.34\textwidth]{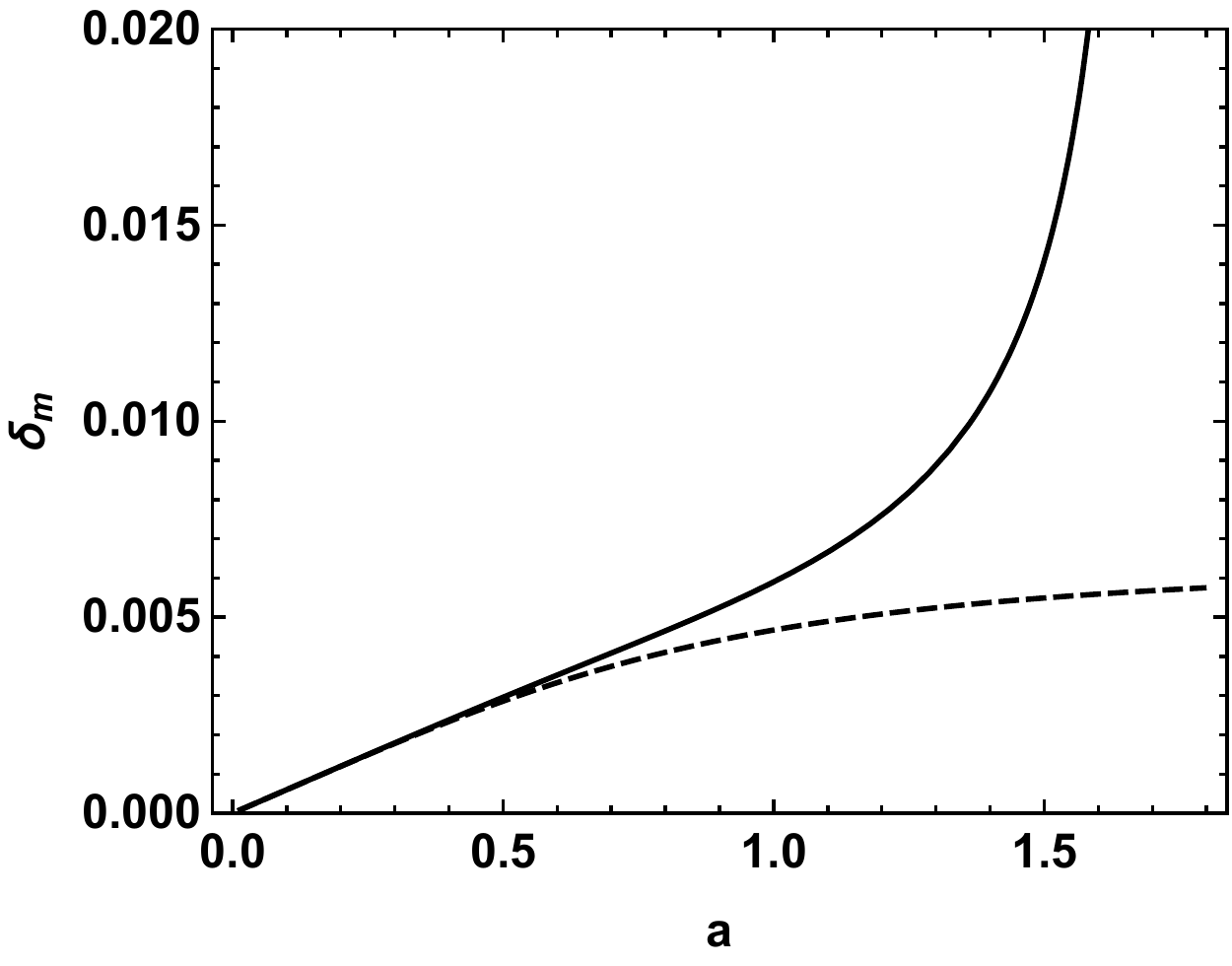} \hspace{.4in} \includegraphics[width=0.34\textwidth]{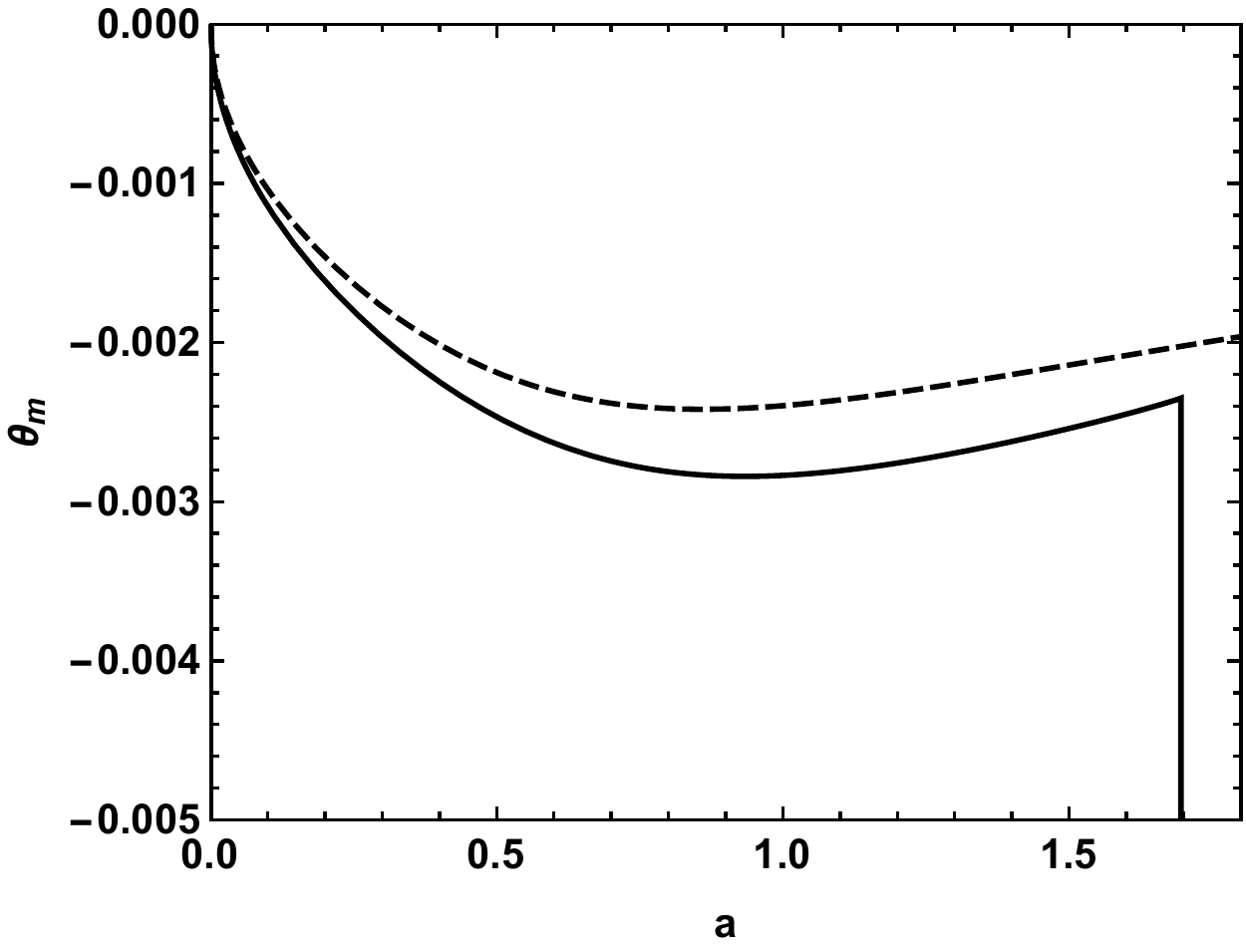}}
\centerline{\includegraphics[width=0.34\textwidth]{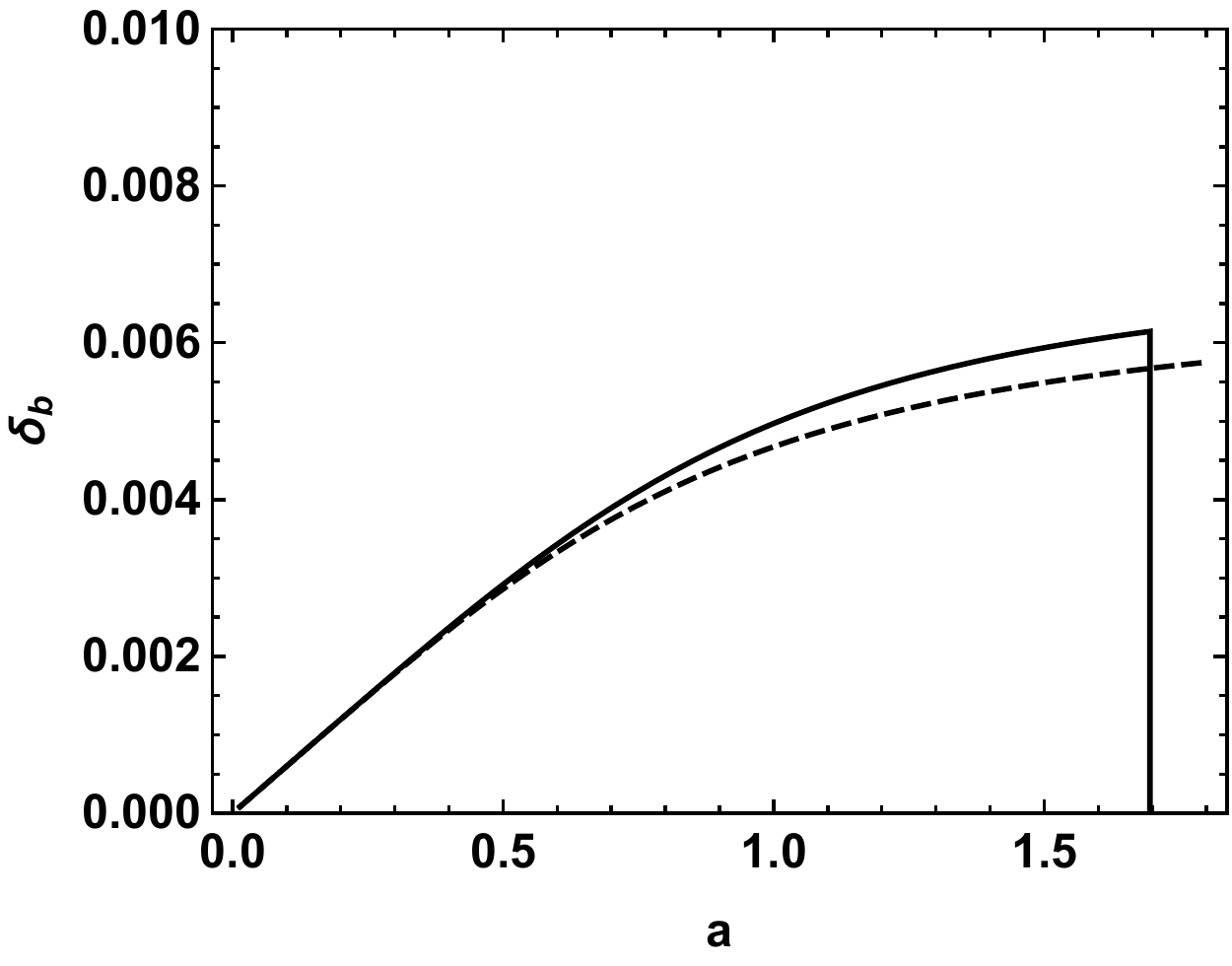} \hspace{.4in} \includegraphics[width=0.34\textwidth]{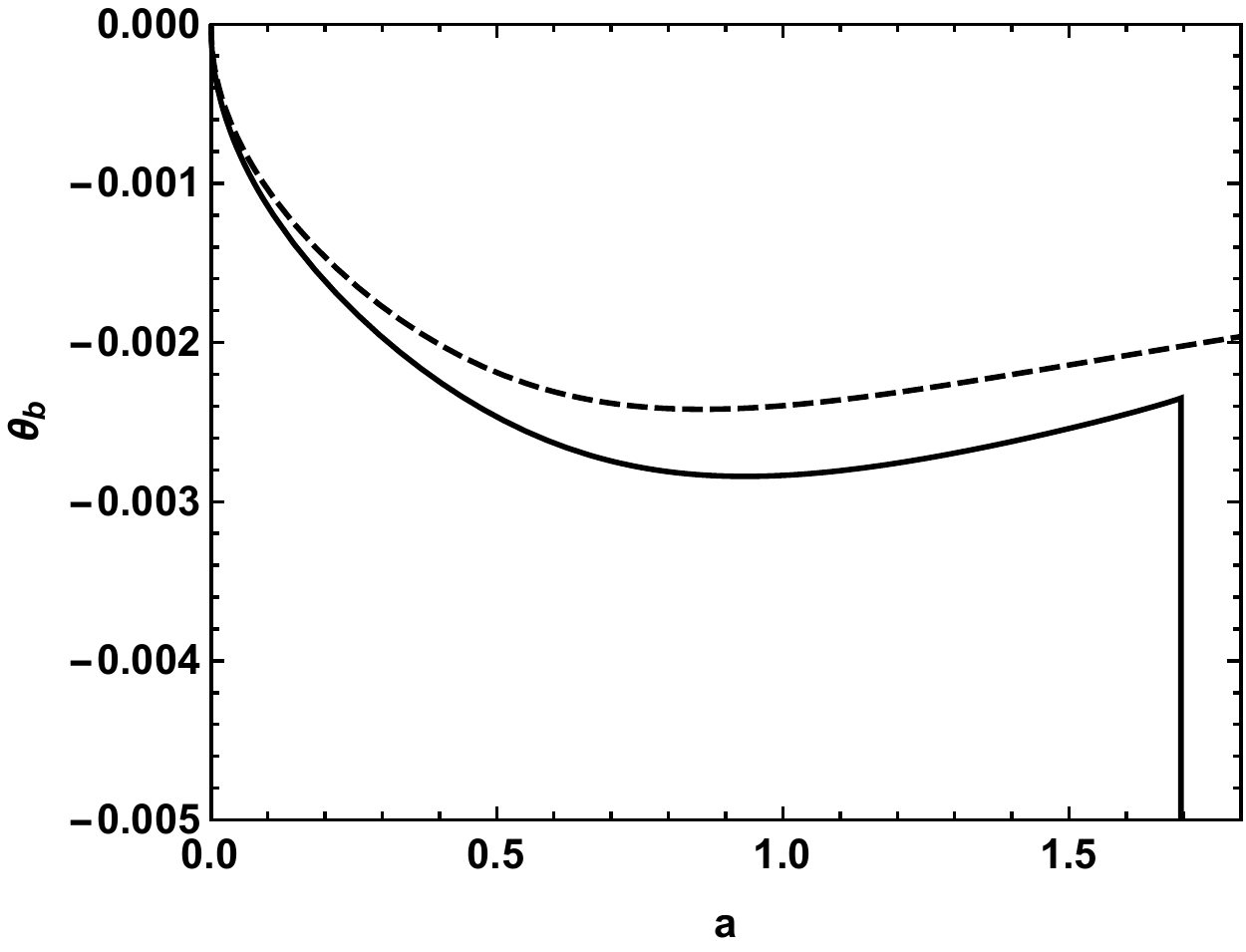}}
\end{center}
\caption{Perturbative solutions in subhorizon limit as functions of the scale factor. \textbf{Top left panel:} Pressureless matter density contrast. \textbf{Top right panel:} Pressureless matter velocity potential. \textbf{Bottom left panel:} Baryon density contrast. \textbf{Bottom right panel:} Baryon velocity potential. In all panels solid lines correspond to $\xi = -0.1$, $\omega_x = -1$ and $\Omega_{x0} = 0.7$, and dashed lines correspond to the $\Lambda$CDM model with the same $\Omega_{x0}$.}
\label{delta}
\end{figure*}

In Fig. 3 we plot the evolution of the total matter/baryon density contrasts and velocity potentials, that show an instability at $a \approx 1.7$, when the matter density becomes negative (for $\xi = -0.1$ and $\Omega_{x0}=0.7$). As can be seen in the top left panel, when $\rho_m$ reaches values arbitrarily close to zero, $\delta_{m}$ diverges. On the other hand, the bottom panels show that this instability propagates, via $\delta \rho_{m}$, to the baryonic component, even though that component does not violate the WEC and remains positive for $a>1.7$. In order to solve the closed system (\ref{matter1})-(\ref{barions2}), we have used $\xi = -0.1$, $\omega_x = -1$ and $\Omega_{x0} = 0.7$, besides the initial conditions $\delta_{m} = \delta_{b} = 10^{-5}$ and $\theta_{m} = \theta_{b} = 0$ at $a = 10^{-3}$.

As we have shown, for $\omega_x = -1$ the DE component is smooth on sub-horizon scales. This is not true if $\omega_x > -1$, when DE clusters and contributes to clustering matter \cite{Carneiro:2014uua}. Therefore, one may argue that the weak energy condition must be satisfied by the total clustering energy, not by pressureless matter alone. In order to verify this possibility, let us decompose the dark energy as
\begin{eqnarray}
\rho_x &=& \rho_{\Lambda} + \rho_{m'}, \label{rhoxdecomp}\\
p_x &=& \omega_x \rho_x = -\rho_{\Lambda},
\end{eqnarray}
where $\rho_{m'}$ is pressureless and $\rho_{\Lambda}$ has EoS parameter $-1$. The total clustering matter (including the fluctuating DE part) is given by
\begin{equation}
\rho_c = \rho_m + \rho_{m'} = \rho_m + (1+\omega_x) \rho_x.
\end{equation}
From (\ref{4})-(\ref{3}) we then have
\begin{eqnarray}
\rho_{\Lambda} &=& -\omega_x \rho_{x0}\, a^{-3(\xi+\omega_x+1)},\\
\rho_c &=& C a^{-3} + \left[ (1+\omega_x) - \left( \frac{\xi}{\xi+\omega_x} \right) \right] \rho_{x0}\, a^{-3(\xi+\omega_x+1)}.
\end{eqnarray}
It is easy to verify that $\rho_c$ is positive definite if, and only if,
\begin{equation} \label{ineq}
\omega_x + \xi + 1 \geq 0.
\end{equation}
When this inequality is saturated, the model is equivalent to a $\Lambda$CDM model. It is violated, in particular, by the best-fit values obtained from current observations \cite{DiValentino:2019ffd}.

With the above decomposition, equations (\ref{DE})-(\ref{materia}) can be rewritten as
\begin{equation}
\dot{\rho}_c + 3 H \rho_c = - \dot{\rho}_{\Lambda} = Q_c,
\end{equation}
with
\begin{equation}
Q_c = 3 (\omega_x + \xi + 1) H \rho_{\Lambda}.
\end{equation}
If (\ref{ineq}) is satisfied, the energy flux is from $\rho_{\Lambda}$ to $\rho_c$.
With this interaction term, the sub-horizon perturbation equations for the clustering matter are the same as (\ref{matter1})-(\ref{matter2}),
\begin{eqnarray}
\theta'_{c}+\mathcal{H}\theta_{c}+\frac{a^2}{2} \rho_{c} \delta_{c}=0,\\
\delta'_{c}+\theta_{c}=-\frac{aQ_c}{\rho_{c}} \delta_{c},
\end{eqnarray}
with $\delta_{\Lambda} \ll \delta_c$ and $\delta Q_c \approx 0$. Under condition (\ref{ineq}), there is no instability. Note that the decomposition defined by Eq.~\eqref{rhoxdecomp} was introduced as part of a complementary discussion on the weak energy condition analysis. This approach must be considered only in this context, and there is no relation with the parameter selection presented in Sec.~\ref{paramselec}.

To conclude this analysis, a discussion on the use of the comoving synchronous gauge is in order. In the synchronous gauge the balance equations are given by
\begin{equation}\label{11}
\dot{\theta}_m+2H\theta_m=\frac{Q}{\rho_m}(\theta-\theta_m)+\frac{\nabla_i\delta\bar Q^i}{\rho_m},
\end{equation}
\begin{equation}\label{12}
\dot{\theta}_x+\bigg[\frac{\dot\omega_x}{1+\omega_x}-\frac{Q}{\rho_x}-H(3\omega_x-2)\bigg]\theta_x=\frac{1}{\rho_x(1+\omega_x)}\bigg[\frac{k^2}{a^2}\delta p_x-Q\theta-\nabla_i\delta\bar Q^i\bigg],
\end{equation}
\begin{equation}\label{13}
\dot\delta_m+\theta_m-\frac{\dot h}{2}=-\frac{Q}{\rho_m}\delta_m+\frac{\delta Q}{\rho_m},
\end{equation}
\begin{equation}\label{14}
\dot\delta_x+3H\bigg[\frac{\delta p_x}{\delta\rho_x}-\omega_x\bigg]\delta_x+(1+\omega_x)\theta_x-(1+\omega_x)\frac{\dot h}{2}=\frac{Q}{\rho_x}\delta_x-\frac{\delta Q}{\rho_x}.
\end{equation}
For the metric potential, we obtain from the Einstein's equations 
\begin{equation}\label{15}
\ddot h+2H\dot h=\rho_m\delta_m+\rho_x\delta_x+3\delta p_x.
\end{equation}
In the case $\omega_x = -1$ we can assume, as above, that matter follows geodesics, that is, we can set $\delta \bar{Q}^i = 0$, $\theta = \theta_m = 0$, and (\ref{11}) is identically satisfied. From (\ref{12}) we see that $\delta \rho_x = -\delta p_x = 0$ (while $\theta_x$ remains undefined). Finally, from (\ref{14}) we have $\delta Q = 0$ and our system is reduced to
\begin{eqnarray}\label{comovel1}
\dot\delta_m-\frac{\dot h}{2}=-\frac{Q}{\rho_m}\delta_m,\\\label{comovel2}
\ddot h+2H\dot h=\rho_m\delta_m.
\end{eqnarray}
Systems (\ref{matter1})-(\ref{matter2}) and (\ref{comovel1})-(\ref{comovel2}) lead to the same second order equation for the matter contrast,
\begin{equation}
    \ddot{\delta}_m + (2H + \Gamma) \dot{\delta}_m + (2H\Gamma + \dot{\Gamma}) \delta_m = \frac{\rho_m \delta_m}{2},
\end{equation}
where we have introduced the rate of matter creation $\Gamma = Q/\rho_m$. It is easy to check that, for $\omega_x \neq -1$, this is not generally possible. Dark energy is perturbed, there is momentum transfer in the matter rest frame, matter does not follow geodesics, which means that it is not comoving with synchronous observers.

\section{Impact of the WEC prior on parameter estimation}
\label{paramselec}

We shall now investigate the impact of the physical or WEC prior, $\xi \geq 0$,  
on the model parameter estimation. 
With and without the inclusion of such a prior we perform a Bayesian statistical analysis considering two different data sets. First, we use the full CMB data from Planck alone, which contains information from temperature and polarization maps and the lensing reconstruction, Planck~(TT,TE,EE+lowE+lensing) \cite{Aghanim:2019ame}. Second we combine the Planck data with the current sample of the Pantheon catalog of type Ia Supernovae (SNe Ia) \cite{Jones:2017udy,Scolnic:2017caz}\footnote{The Pantheon data can be downloaded from \href{https://github.com/dscolnic/Pantheon}{github.com/dscolnic/Pantheon}. For CMB analysis, all Planck likelihood codes and data can be obtained at \href{http://pla.esac.esa.int/pla/}{pla.esac.esa.int/pla}.}.

We choose not to use the Baryonic Acoustic Oscillations (BAO) data 
as there is no evidence that the fiducial model used in the BAO peak extraction would not bias our analysis, contrary to the usual $\Lambda$CDM parameters~\cite{Carter:2019ulk}. Furthermore, we emphasize that the parameter estimation itself is not the main goal of the paper, but rather to assess how the inclusion of the WEC prior affects the parameter constraints.

In order to perform the parameter selection we make use of a suitable modified version of the Boltzmann code {\sc class}~\cite{Blas:2011rf} for the studied interacting model, combined with {\sc MontePython}~\cite{Audren:2012wb,Brinckmann:2018cvx} for running the MCMC process. The cosmological parameters considered in our analysis are the six usual ones plus the interaction parameter, i.e., $\theta=\{ \omega_{b }, \omega_{c }, \ln (10^{10}A_{s}), n_{s }, \tau_{reio }, H_{0}, \xi \}$, with the dark energy EoS parameter $w_{x}$ = -1. The results of our analysis are presented in Tab.~\ref{tab}.
\begin{table}[]
\begin{tabular}{l|l|l|l|l}
\hline \hline 
                     & \multicolumn{2}{c|}{Planck}                                          & \multicolumn{2}{c}{Planck+Pantheon}                           \\ \hline
Parameter            & \multicolumn{1}{c|}{No prior}        & \multicolumn{1}{c|}{WEC prior} & \multicolumn{1}{c|}{No prior} & \multicolumn{1}{c}{WEC prior} \\ \hline
$100~\omega_{b }$    & ($2.234$) $2.242_{-0.017}^{+0.015}$      & ($2.236$) $2.243_{-0.012}^{+0.018}$       & \multicolumn{1}{l|}{($2.223$) $2.236_{-0.014}^{+0.016}$}      & ($2.240$) $2.239_{-0.015}^{+0.014}$ \\
$\omega_{c }$        & ($0.1197$) $0.1194_{-0.0014}^{+0.0014}$  & ($0.1195$) $0.1192_{-0.0012}^{+0.0014}$   & \multicolumn{1}{l|}{($0.1210$) $0.1202_{-0.0013}^{+0.0012}$}   & ($0.1203$) $0.1197_{-0.0012}^{+0.0012}$ \\
$\ln (10^{10}A_{s})$ & ($3.026$) $3.037_{-0.016}^{+0.016}$      & ($3.037$) $3.034_{-0.013}^{+0.016}$       & \multicolumn{1}{l|}{($3.044$) $3.047_{-0.015}^{+0.014}$}      & ($3.046$) $3.046_{-0.014}^{+0.016}$ \\
$n_{s }$             & ($0.967$) $0.967_{-0.005}^{+0.005}$      & ($0.967$) $0.967_{-0.005}^{+0.005}$       & \multicolumn{1}{l|}{($0.961$) $0.964_{-0.004}^{+0.004}$}      & ($0.962$) $0.966_{-0.005}^{+0.004}$ \\
$\tau_{reio }$       & ($0.047$) $0.052_{-0.008}^{+0.008}$      & ($0.053$) $0.051_{-0.007}^{+0.008}$       & \multicolumn{1}{l|}{($0.053$) $0.055_{-0.008}^{+0.007}$}      & ($0.054$) $0.056_{-0.008}^{+0.008}$ \\
$H_{0}$              & ($63.12$) $63.75_{-3.0}^{+3.3}$          & ($65.09$) $62.52_{-2.1}^{+3.6}$           & \multicolumn{1}{l|}{($67.58$) $67.73_{-1.0}^{+1.0}$}                                   & ($67.35$) $66.8_{-0.63}^{+0.73}$ \\
$\xi$                & ($0.15$) $0.13_{-0.12}^{+0.11}$          & ($0.09$) $0.18_{-0.12}^{+0.11}$           & \multicolumn{1}{l|}{($-0.02$) $-0.01_{-0.04}^{+0.04}$}                                 & ($0.00$) $0.02_{-0.02}^{+0.01}$ \\ \hline
$\Omega_{m0}$        & ($0.36$) $0.35_{-0.04}^{+0.03}$          & ($0.33$) $0.36_{-0.04}^{+0.02}$           & \multicolumn{1}{l|}{($0.31$) $0.31_{-0.01}^{+0.01}$}                                   & ($0.31$) $0.32_{-0.01}^{+0.01}$ \\
$\sigma_{8}$         & ($0.74$) $0.75_{-0.06}^{+0.05}$          & ($0.77$) $0.73_{-0.04}^{+0.05}$           & \multicolumn{1}{l|}{($0.82$) $0.82_{-0.02}^{+0.02}$}                                   & ($0.81$) $0.80_{-0.01}^{+0.02}$ \\ \hline
$\chi^{2}_{min}$     & \multicolumn{1}{c|}{$2777.24$}           & \multicolumn{1}{c|}{$2778.56$}            & \multicolumn{1}{c|}{$3804.80$}                                                         & \multicolumn{1}{c}{$3806.04$} \\ \hline\hline
\end{tabular}
\caption{Results of the statistical analysis. The result is presented with the best-fit (in parenthesis) and the mean$\pm 1\sigma$ CL.}
\label{tab}
\end{table}

\begin{figure*}
\centerline{\includegraphics[width=0.315\textwidth]{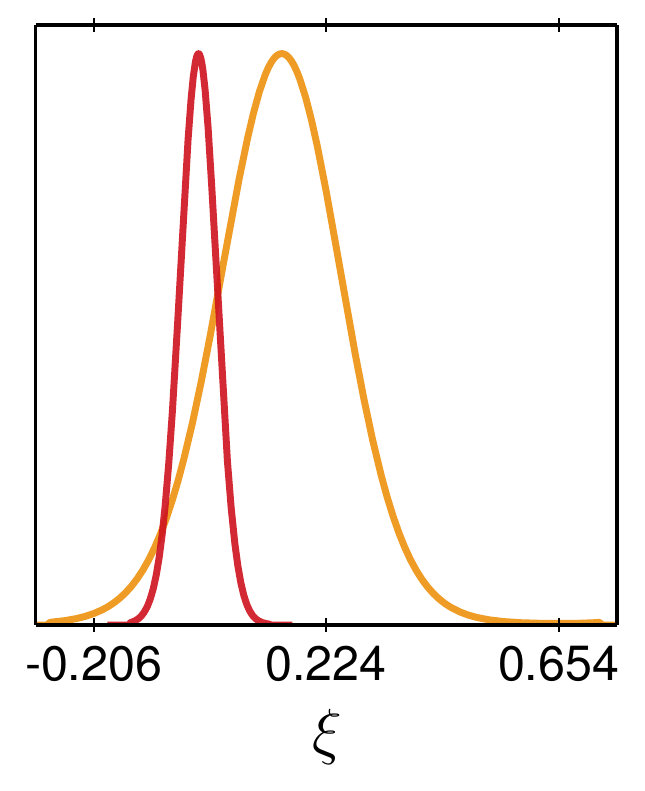} 
\includegraphics[width=0.34\textwidth]{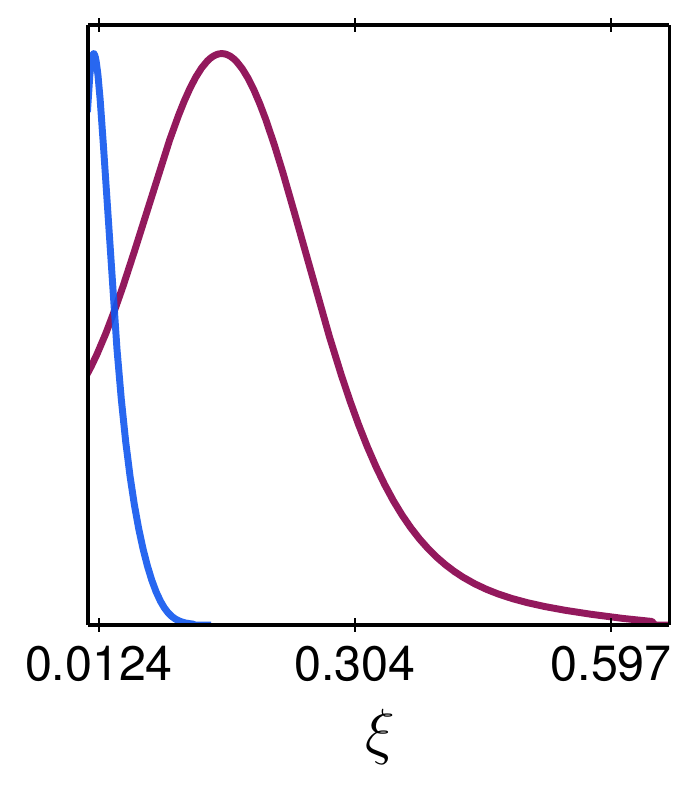} 
\includegraphics[width=0.31\textwidth]{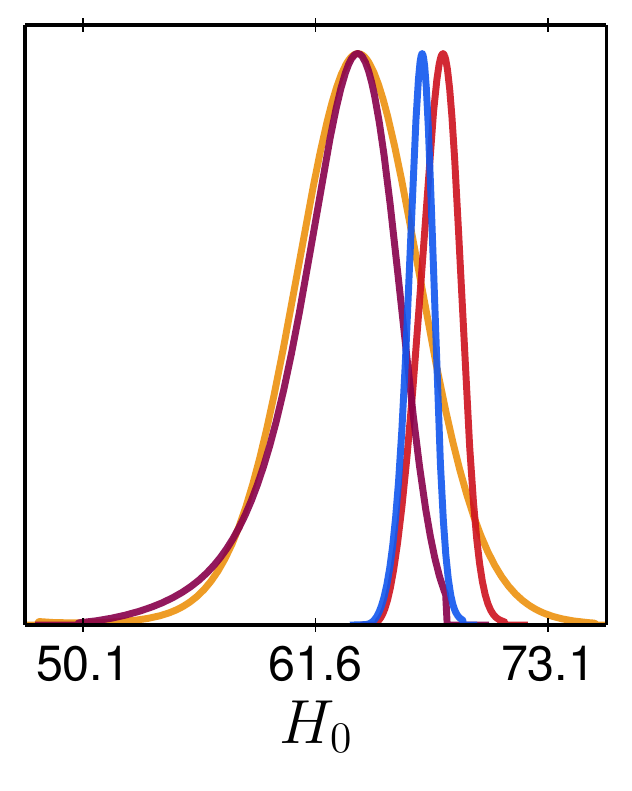}}
 \caption{Posteriors for the interaction parameter without (left) and with (centre) the WEC prior. The panel on the right shows the posteriors for $H_{0}$. \textbf{Orange lines: }Planck (no prior). \textbf{Red lines: }Planck+Pantheon (no prior). \textbf{Purple lines: }Planck (WEC prior). \textbf{Blue lines: }Planck+Pantheon (WEC prior).}
  \label{figxi}
 \end{figure*}

Figure~\ref{figxi}  shows the posteriors obtained for the interaction parameter $\xi$ for the cases where no prior (left panel) and the WEC prior (centre panel) are used. As can be seen, the $\Lambda$CDM limit ($\xi=0$) is obtained within $1\sigma$ (C.L.) in all cases. 
For the case with no prior, when only Planck data is used, the best-fit and mean value of $\xi$ are both positive, which avoid the WEC violation discussed earlier. On the other hand, when the Pantheon data is added to the analysis, the best-fit and mean value of $\xi$ are slightly smaller than zero, which means that, inevitably, the WEC will be violated in the future. Notwithstanding, when the WEC prior is taken into account, the analysis for $\xi$ is not drastically affected. When only the Planck data is employed the best-fit and mean values for $\xi$ are also positive, in agreement with the analysis with no prior. For the combined Planck+Pantheon data set, the constraint on $\xi$ has the $\Lambda$CDM limit as the best-fit, with an upper-bound limit compatible with the previous analysis with the WEC prior. It is worth mentioning that the interval of $\xi$ derived from the past WEC condition (Eq.~(\ref{prior1})) agrees with all analysis, which amounts to saying that the impact on the parameter estimation comes from the future WEC condition, i.e., the positivity of the interacting parameter $\xi$ defined in Eq.~\eqref{prior2}.

The posteriors for $H_{0}$ are shown in the right panel of Fig.~\ref{figxi}. Both analyses, with and without the WEC prior, have weak constraints on the Hubble constant when we consider only the Planck data, with best-fit and mean around $H_{0}\approx 64\ {\rm km}\ {\rm s}^{-1}\ {\rm Mpc}^{-1}$. The wide error bars of $H_{0}$ in these cases evidence the necessity of adding more data to the analysis. Adding the SNe Ia data from Pantheon, the studied model seems to alleviate the $H_{0}$ tension, which is related to the negative value of the interacting parameter $\xi$. The inclusion of the WEC prior 
moves the constraints towards smaller values of $H_{0}$, which increases the well-known tension with local measurements of the current expansion rate~\cite{Riess:2019cxk}.

\begin{figure*}
\centerline{\includegraphics[width=0.8\textwidth]{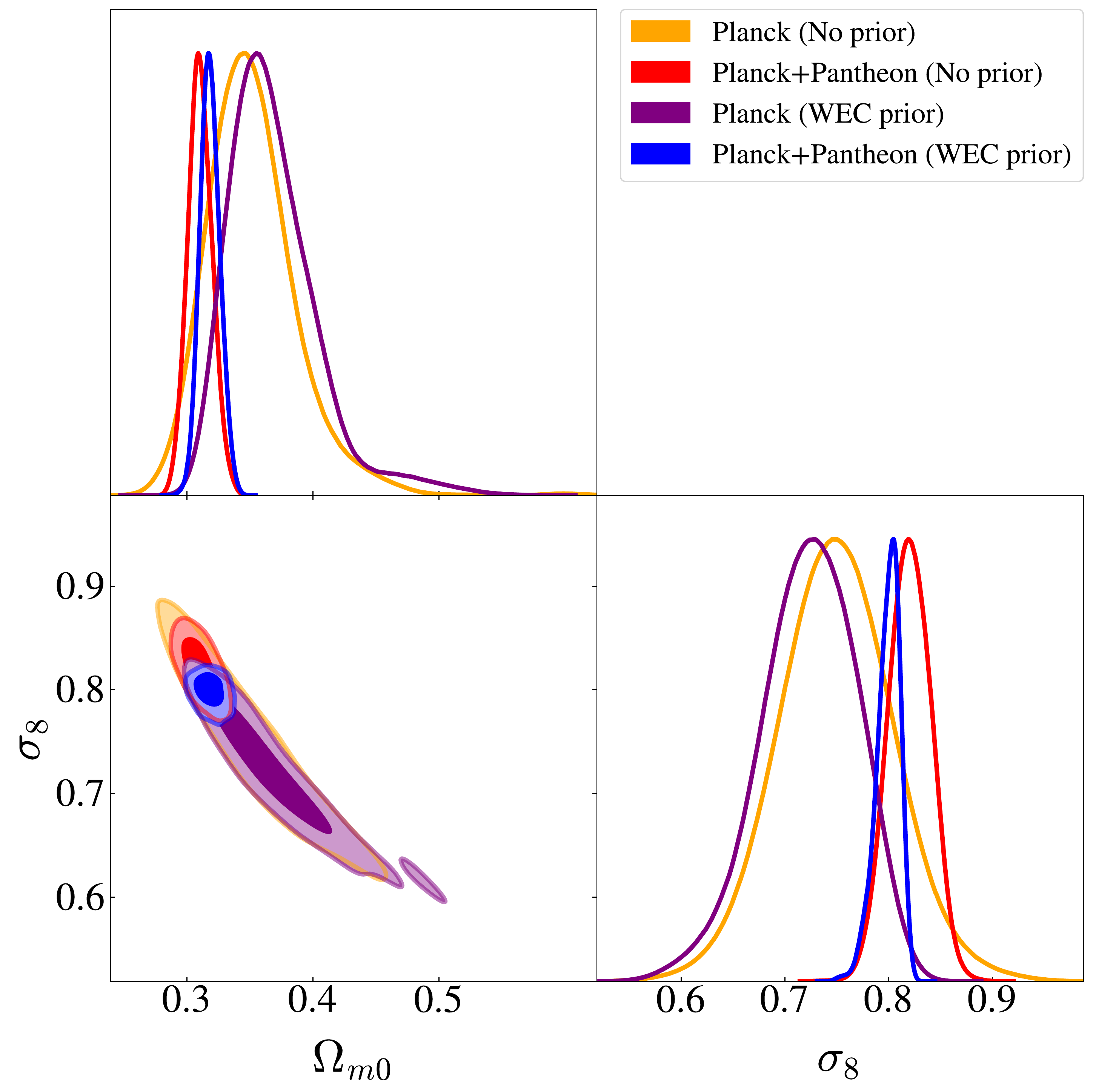}}
 \caption{Corner plot for $\Omega_{m0}$ and $\sigma_{8}$.}
  \label{figoms8}
 \end{figure*}

Finally, we show the corner plot for the plane $\Omega_{m0}$ - $\sigma_{8}$ in Fig.~\ref{figoms8}. Similarly to the $H_{0}$ analysis, in the case where only the Planck data is employed, we have weak constrains on $\Omega_{m0}$ and $\sigma_{8}$. In particular, the predictions for $\Omega_{m0}$ and $\sigma_{8}$ in both analyses are about $0.35$ and $0.74$ respectively. On the other hand, when the Pantheon data is combined with Planck data, the constraints on $\Omega_{m0}$ and $\sigma_{8}$ are considerably improved. Also in this case, the inclusion of the WEC prior alters the parameter constraints, with a slight preference for smaller values of $\sigma_8$, which somehow seems to alleviate the $\sigma_{8}$ tension between the Planck primary CMB results and estimates from cosmic shear surveys~\cite{Joudaki:2017zdt}. For example, a naive comparison between our estimates of the quantity $S_8 = \sigma_8(\Omega_m/0.3)^{1/2}$ with and without the WEC prior with the one provided by the KiDS weak lensing survey~\cite{Joudaki:2017zdt} shows a difference of $1.5\sigma$ and $1.64\sigma$, respectively. An even better agreement ($\lesssim1\sigma$) is obtained from a comparison using the current clustering and lensing data from the Dark Energy Survey \cite{Troxel:2017xyo}.

\section{Conclusions}

Interacting DE cosmologies are capable of providing a good description of the Universe evolution and constitute an alternative to the standard cosmological model~\cite{Benetti:2019lxu}. As recently reported in \cite{DiValentino:2019ffd}, the current observational data seem to favor the former class of models with an interaction term of the type $Q = 3\xi H \rho_{x}$. In particular, for negative values of the interacting parameter {\bf ($3\xi \simeq -0.5$)}, that study showed that this model is able to provide a solution for the widely discussed $H_0$ and $\sigma_8$-tension problems.

In this paper, we have investigated the theoretical consistency of this class of cosmologies and shown that for negative values of $\xi$, which physically corresponds to a transfer of energy from dark matter to dark energy, this particular model predicts an eventual violation of the WEC ($\rho \geq 0$) for the dark matter density, which results in the instabilities of the matter perturbations discussed in Sec. III. For completeness, we have also discussed the impact of the physical prior, $\xi \geq 0$, on the model parameter estimation through a statistical analysis of the latest CMB and type Ia Supernovae data. The results show that the model predictions are in good agreement with current estimates of $\sigma_8$ but far from providing a possible solution for the $H_{0}$-tension problem.

\section*{Acknowledgements}

RvM acknowledges support from CNPq and from the Federal Commission for Scholarships for Foreign Students for the Swiss Government Excellence Scholarship (ESKAS No. 2018.0443) for the academic year 2018-2019. SC is partially supported by CNPq with grant No. 307467/2017-1. JSA acknowledges support from CNPq (grants No. 310790/2014-0 and 400471/2014-0) and FAPERJ (grant No. E-26/203.024/2017).

\appendix

\section{Corner plots}

\begin{figure*}
\centerline{\includegraphics[width=\textwidth]{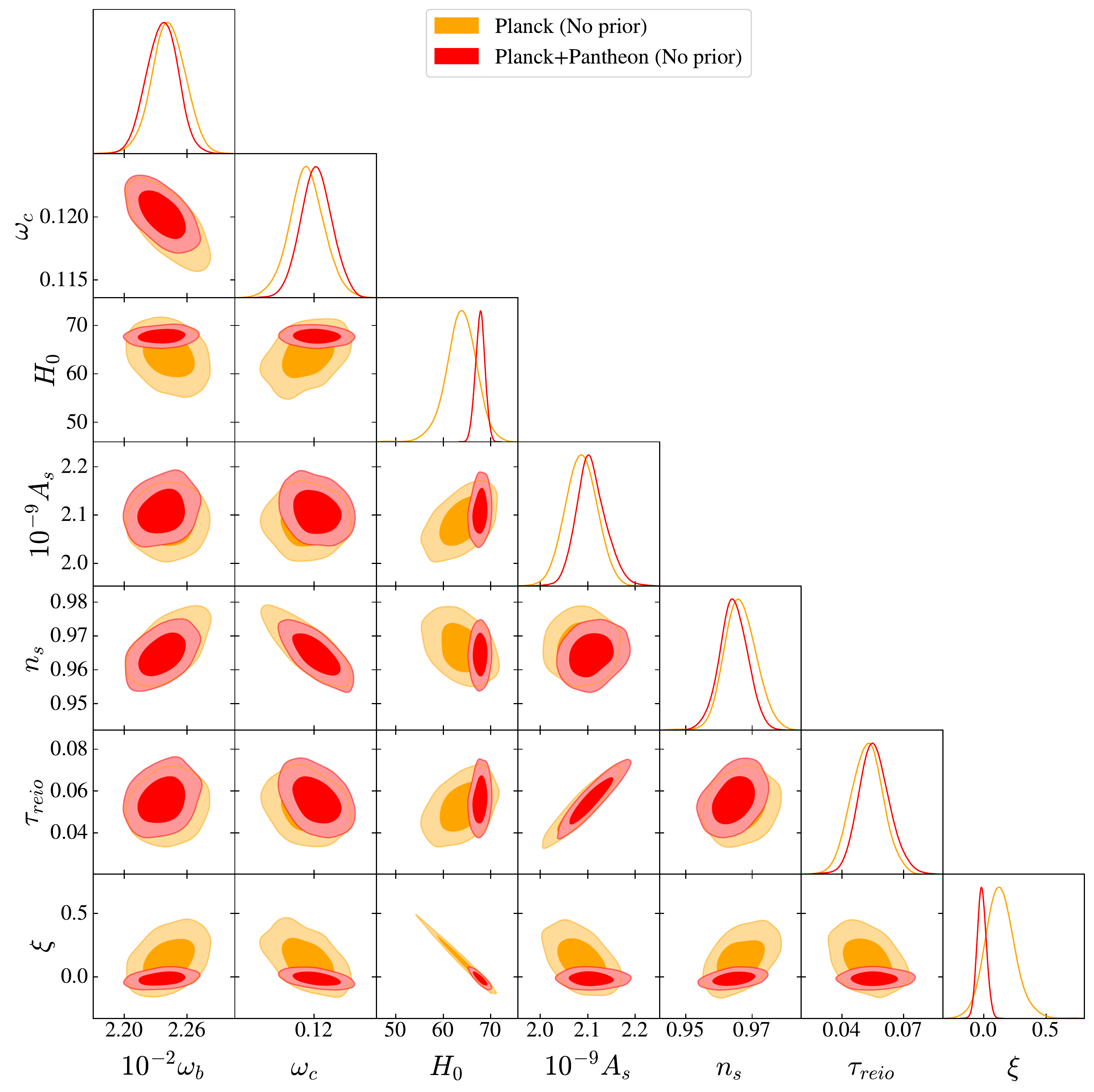}}
 \caption{Triangle plot with the free cosmological parameters for the analyses where the WEC prior is not taken into account using Planck and Planck+Pantheon data.}
  \label{triangle}
 \end{figure*}

\begin{figure*}
\centerline{\includegraphics[width=\textwidth]{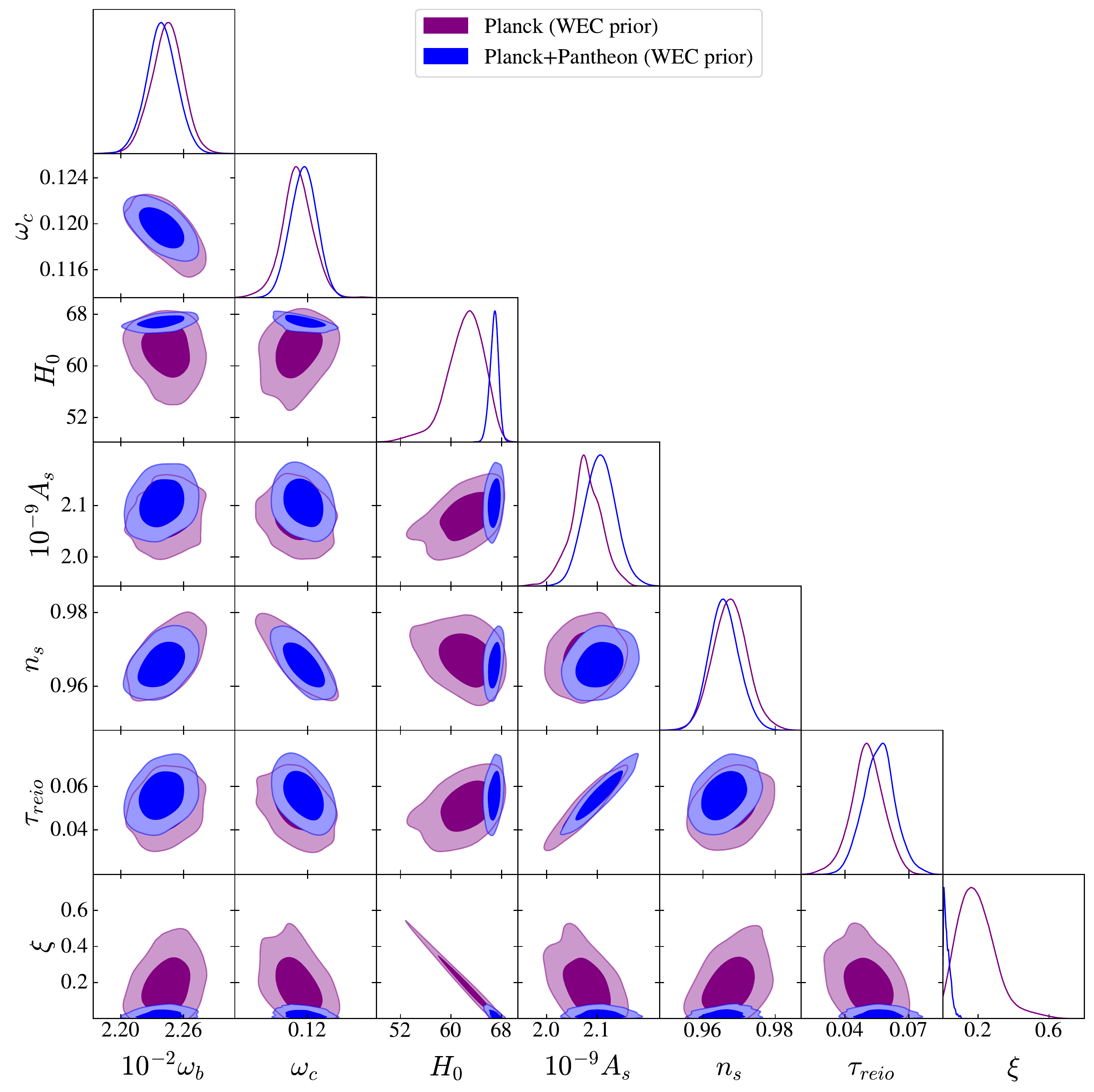}}
 \caption{Triangle plot with the free cosmological parameters for the analyses where the WEC prior is taken into account using Planck and Planck+Pantheon data.}
  \label{trianglewec}
 \end{figure*}

For the sake of completeness, we show the resulting triangle plots with all free cosmological parameters. These plots are particularly useful because they illustrate the correlation between all parameters. In Fig.~\ref{triangle} we show the result for the analyses with no prior, while Fig.~\ref{trianglewec} shows the result for the case where the WEC prior is taken into account.

\bibliography{mybib}

\end{document}